\documentclass[apj]{emulateapj}
\usepackage{natbib}
\usepackage{amsmath}
\usepackage{bm}

\shorttitle{Harmonic in-painting of CMB map}
\shortauthors{Jaiseung Kim and et al.}

\begin{document}
\title{Harmonic in-painting of CMB sky by constrained Gaussian realization}
\author{Jaiseung Kim and Pavel Naselsky}
\affil{Niels Bohr Institute \& Discovery Center, Blegdamsvej 17, DK-2100 Copenhagen, Denmark}
\email{jkim@nbi.dk}
\author{Nazzareno Mandolesi}
\affil{INAF/IASF, Istituto di Astroﬁsica Spaziale e Fisica Cosmica di Bologna, Istituto Nazionale di Astroﬁsica, via Gobetti 101, I-40129 Bologna, Italy}

\submitted{Submitted to the Astrophysical Journal Letter} 

\begin{abstract}
The presence of astrophysical emissions between the last scattering surface and our vantage point requires us to apply a foreground mask on CMB sky map, leading to large cut around the Galactic equator and numerous holes.
Since many CMB analysis, in particular on the largest angular scales, may be performed on a whole sky map in a more straightforward and reliable manner, it is of utmost importance to develop an efficient method to fill in the masked pixels in a way compliant with the expected statistical properties and the unmasked pixels.
In this letter, we consider Monte-Carlo simulation of constrained Gaussian field and derive it for CMB anisotropy in harmonic space, where a feasible implementation is possible with good approximation. We applied our method to simulated data, which shows our method produces a plausible whole-sky map, given the unmasked pixels and a theoretical expectation. Subsequently, we applied our method to the WMAP foreground-reduced maps and investigated the anomalous alignment between quadrupole and octupole components. 
From our investigation, we find the alignment in the foreground-reduced maps is even higher than the ILC map. We also find the V band map has higher alignment than other bands, despite the expectation that the V band map has less foreground contamination than other bands. Therefore, we find it hard to attribute the alignment to residual foregrounds.
Our method will be complementary to other efforts on in-painting or reconstructing the masked CMB data, and of great use to Planck surveyor and future missions.
\end{abstract}


\section{Introduction}
\label{intro}
There exist several astrophysical emission sources between the last scattering surface and our vantage point.
Due to the contamination from the `foregrounds', we need to apply proper masking on microwave sky maps, which leads to cut of varying width around the Galactic equator and numerous holes. Since many CMB analysis, in particular on the largest angular scales, may be performed on a whole-sky map in a more straightforward and reliable manner, there have been several efforts to reconstruct a whole-sky map from incomplete sky data \citep{lowl_WMAP1,PE_low,lowl_foreground,lowl_recon}.
However, the fidelity of reconstruction is limited, because it is not possible to reliably reconstruct harmonics modes mainly confined to the Galactic cuts \citep{PE_low}.
Therefore, there have been active attempts to fill in the missing information in CMB sky data with a priori \citep{in-painting_sparse,in-painting_inoue,in-painting_Bucher}. 

Historically, the act of recovering damaged parts of valuable paintings by a skilled restoration artist is called `in-painting'.
In digital imaging, there are various in-painting methods \citep{in-painting_Masnou,in-painting_Ballester,in-painting_Bertalmio,in-painting_Rane}. 
While these methods work well for images of periodic or predictable patterns, they may not be suitable for CMB data, which have random Gaussian nature.
On the other hand, there have been works on generating constrained Gaussian fields, which have been used in the study of large-scale structures \citep{Constrained_Gaussian_Path,Constrained_Gaussian,Constrained_Gaussian_Simple}.
However, it is not feasible for the pixel data of the WMAP or Planck surveyor, which amounts to millions of pixels or more.
In this letter, we are going to implement the method for CMB anisotropy in harmonic space, where the computational load may be significantly reduced with good approximation. After demonstrating it with simulated data, we are going to apply our method to the WMAP foreground-reduced maps and investigate the well-known anomaly associated with the quadrupole and octupole component. Throughout this letter, we will use the term `in-painting' to denote our described procedure.

The outline of this paper is as follows.
In Section \ref{CMB}, we briefly discuss CMB anisotropy in harmonic space and the effect of incomplete sky coverage. 
In Section \ref{harmonic}, we discuss simulation of constrained Gaussian field and its implementation for CMB anisotropy in harmonic space.
In Section \ref{simulation}, we apply our method to simulated data and present the result.
In Section \ref{wmap}, we apply our method to masked WMAP data and investigate the multipole vector alignment between the quadrupole and octupole.
In Section \ref{discussion}, we summarize our work.

\section{CMB anisotropy in harmonic space}
\label{CMB}
CMB anisotropy over a whole-sky is conveniently decomposed in terms of spherical harmonics: 
\begin{eqnarray}
T(\hat {\mathbf n})&=&\sum_{lm} a_{lm}\,Y_{lm}(\hat {\mathbf n}),\label{T_expansion}
\end{eqnarray}
where $a_{lm}$ and $Y_{lm}(\theta,\phi)$ are a decomposition coefficient and a spherical harmonic function, and  $\hat {\mathbf n}$ denotes a sky direction.
In most of inflationary models, decomposition coefficients of CMB anisotropy follow the Gaussian distribution of the following statistical properties:
\begin{eqnarray}
\langle a_{lm} \rangle &=& 0,\label{mean_alm}\\
\langle a_{lm}a^*_{l'm'}\rangle &=& \delta_{ll'} \delta_{mm'}\,C_{l}, \label{Cl}
\end{eqnarray}
where $\langle \ldots \rangle$ denotes the average over an ensemble of universes, and
$C_l$ denotes CMB power spectrum.
Accordingly, the CMB anisotropy $T(\theta,\phi)$, which follows Gaussian distribution, have the following angular correlation:
\begin{eqnarray}
\langle T(\hat {\mathbf n})\,\,T(\hat {\mathbf n'}) \rangle = \sum_l \frac{2l+1}{4\pi}\,W_l\,C_l\,P_l(\cos\theta),\label{C}
\end{eqnarray}
where $P_l$ is a Legendre polynomials and $\theta=\cos^{-1}(\hat{\mathbf n}\cdot \mathbf {\hat n'})$.

In the presence of a foreground mask, the spherical coefficients of a masked sky $\tilde a_{lm}$ are related to those of a whole-sky as follows:
\begin{eqnarray}
\tilde a_{l_3m_3}=\sum_{l_2 m_2} F(l_2,m_2,l_3,m_3)\,a_{l_2 m_2}\label{a_LM},
\end{eqnarray}
where
\begin{eqnarray}
F(l_2,m_2,l_3,m_3)&=&(-1)^{m_3}\sqrt{\frac{2l_3+1}{4\pi}}\sum_{l_1 m_1}\sqrt{(2l_1+1)(2l_2+1)}\nonumber\\
&&\times \left(\begin{array}{ccc}l_1&l_2 &l_3\\m_1&m_2&-m_3\end{array}\right)
\left(\begin{array}{ccc}l_1&l_2&l_3\\0&0&0\end{array}\right)w_{l_1m_1},\label{F}
\end{eqnarray}
\begin{eqnarray}
w_{l_1 m_1}&=&\int Y^*_{l_1 m_1}(\theta,\phi)\,W(\theta,\phi)\,d\Omega,\label{wlm}
\end{eqnarray}
and $W(\theta,\phi)$ is the mask function, which is zero inside the mask and one elsewhere, and the terms with the big parenthesis being Wigner 3j symbols.
Using Eq. \ref{Cl} and \ref{a_LM}, we may easily show the following correlation between $a_{lm}$ and $\tilde a_{l'm'}$:
\begin{eqnarray}
\langle  a_{lm}\, \tilde a^*_{l'm'}\rangle=F^*(l,m,l',m')\,\,C_{l},\label{FCl}
\end{eqnarray}
\begin{eqnarray}
\langle  \tilde a_{l'm'}\, \tilde a^*_{l''m''}\rangle=\sum_{lm} F^*(l,m,l',m')\,F(l,m,l'',m'')\,\,C_{l}\nonumber\\\label{FFCl}
\end{eqnarray}

\section{In-painting in harmonic space}
\label{harmonic}
\cite{Constrained_Gaussian_Simple} developed an efficient algorithm on the simulation of constrained Gaussian fields.
According to the work, we may simulate Gaussian random field $f(\mathbf r)$ under constraints $f(\mathbf r_j)$ by the following:
\begin{eqnarray}
f(\mathbf r)=f_{\mathrm{mc}}(\mathbf r) + \sum_{ij}\mathbf b_{i}\,\,\,(\mathbf C^{-1})_{ij}\,(f(\mathbf r_j)-f_{\mathrm{mc}}(\mathbf r_{j})),\nonumber\\\label{cg}
\end{eqnarray}
where
\begin{eqnarray}
\mathbf b_{i}&=&\langle f(\mathbf r)\, f(\mathbf r_i)\rangle,\nonumber\\
\mathbf C_{ij}&=&\langle f(\mathbf r_i)\, f(\mathbf r_j)\rangle,\nonumber
\end{eqnarray}
and the subscript `mc' denotes unconstrained Gaussian Monte-Carlo simulation.
Applying Eq. \ref{cg} to CMB pixel data, we may fill in the masked pixel data with plausible values. The procedure has been extensively studied and discussed by \citep{in-painting_Bucher}. Though in a slight different context, the power spectrum estimation by the Gibbs sampling, which includes generating an underlying CMB map according to the conditional distribution, have some overlapping \citep{Gibbs_global,Gibbs_power}. 

As seen in Eq. \ref{C}, there exist pixel correlation at wide range of separation angles, which makes us to take into account tremendous amount of pixels even for filling in a single pixel. Therefore, we may not readily apply the method to the WMAP or Planck data, which have millions of pixels.
Noting CMB anisotropy in harmonic space (i.e. $a_{lm}$) are expected to follow Gaussian distribution, we may consider implementing the contrained Gaussian realization in harmonic space. In this case, we draw each $a_{lm}$ under the constraints, which are masked spherical harmonic coefficients $\tilde a_{l'm'}$.
Rewriting Eq. \ref{cg} explicitly for spherical harmonic coefficients yields:
\begin{eqnarray}
a_{lm}=a^{\mathrm{mc}}_{lm} + \mathbf b \,\mathbf C^{-1}\,(\tilde {\bm a}-\tilde {\bm a}^{\mathrm{mc}}),\label{cgh}
\end{eqnarray}
where $\tilde {\bm a}$ is a column vector consisting of $\tilde a_{l'm'}$ and
\begin{eqnarray}
\mathbf b&=&\langle a_{lm}\;\tilde {\bm a}^\dagger \rangle,\nonumber\\
\mathbf C&=&\langle  \tilde {\bm a}\,\tilde {\bm a}^\dagger \rangle,\nonumber
\end{eqnarray}
and $\tilde a_{lm}$ is a spherical harmonic coefficient of masked sky (c.f. Eq. \ref{a_LM}) and
$\dagger$ denotes a complex conjugate transpose.
Given Eq. \ref{FCl} and \ref{FFCl}, this process by Eq. \ref{cgh} may seem computationally prohibitive.
However, for an widely used foreground mask such as KQ85, the magnitude of $w_{l_1m_1}$ are significant only at lowest multipoles (i.e. $|w_{l_1\gg1,m_1}| \approx 0$).
For such a mask, we may easily show $F^*(l,m,l',m') \approx 0$ for $|l-l'|\gg 1$, using $|w_{l_1\gg1,m_1}| \approx 0$ and the triangular inequalities of Wigner 3j symbol $|l_3-l_2|\le l_1$ (c.f Eq. \ref{FCl}). It is worth to notice that $F^*(l,m,l',m')$ approaches $\delta_{ll'} \delta_{mm'}$ in the limit of a complete sky coverage (i.e. $w_{l_1>0,m_1}=0$). Using this result with Eq. \ref{FCl}, we may subsequently show $\langle  a_{lm}\, \tilde a^*_{l'm'}\rangle$ for $|l-l'|\gg 1$. In other words, $a_{lm}$ are nearly independent of $\tilde a_{l'm'}$, if the multipole number $l$ and $l'$ differ significantly.
Therefore, for a constrained simulation of $a_{lm}$ by Eq. \ref{cgh}, we need to consider only constraints $\tilde a_{l'm'}$ of multipoles not far away from $a_{lm}$.
With this finding, we may significantly reduce the computational cost involved with Eq. \ref{cgh}, and effectively in-paint a masked CMB sky map.  We may summarize the procedure as follows: First, we generate unconstrained $a^{\mathrm{mc}}_{lm}$ by Monte-Carlo simulation, and then transform them by Eq. \ref{cgh}. From the result of Eq. \ref{cgh}, we synthesize CMB anisotropy map $T(\theta,\phi)$, as necessary. 
Throughout this letter, we will refer to this procedure as `in-painting'.

\section{Application to simulated data}
\label{simulation}
\begin{figure}
\centering
\includegraphics[width=0.4\textwidth]{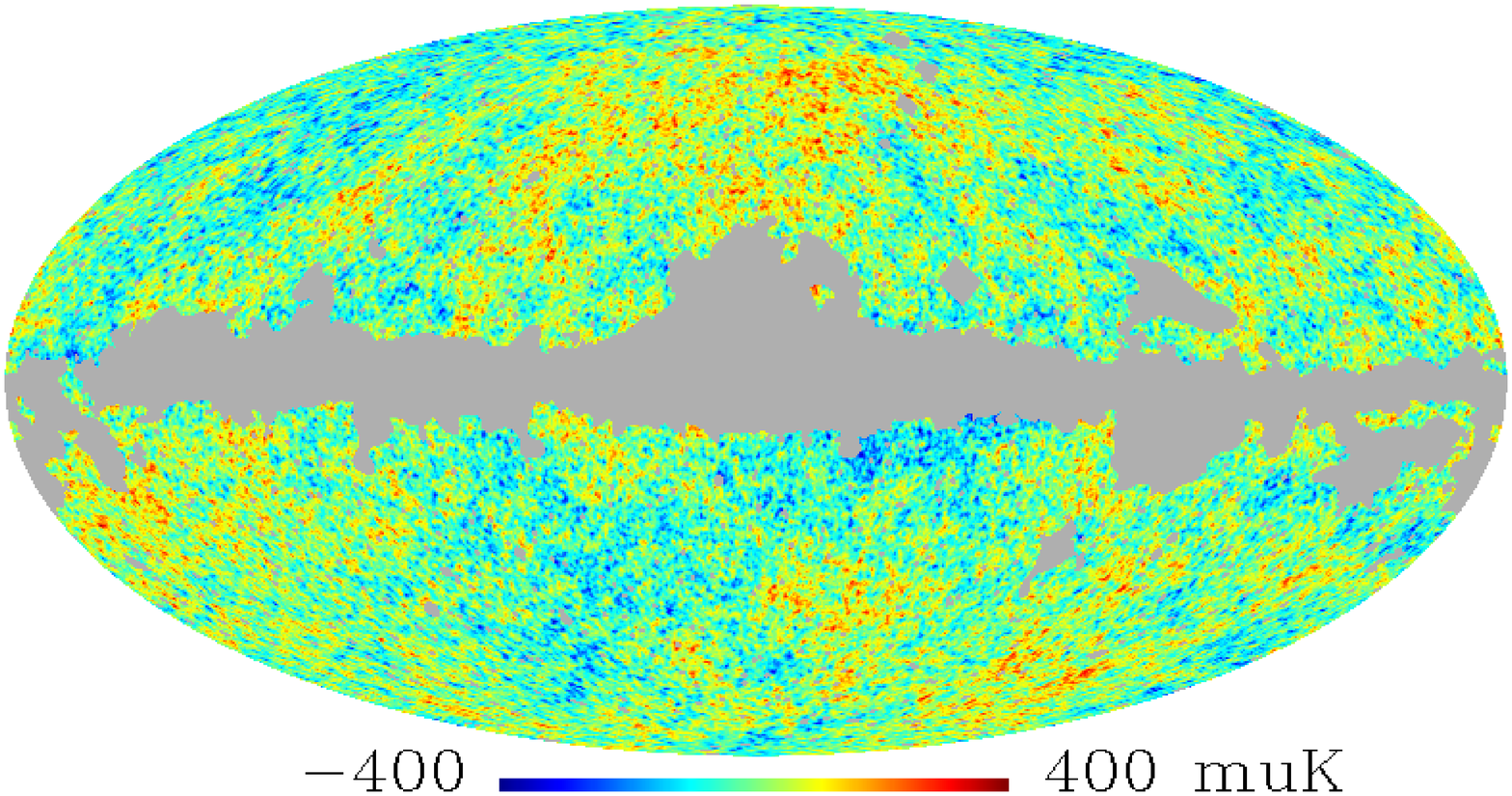}
\includegraphics[width=0.4\textwidth]{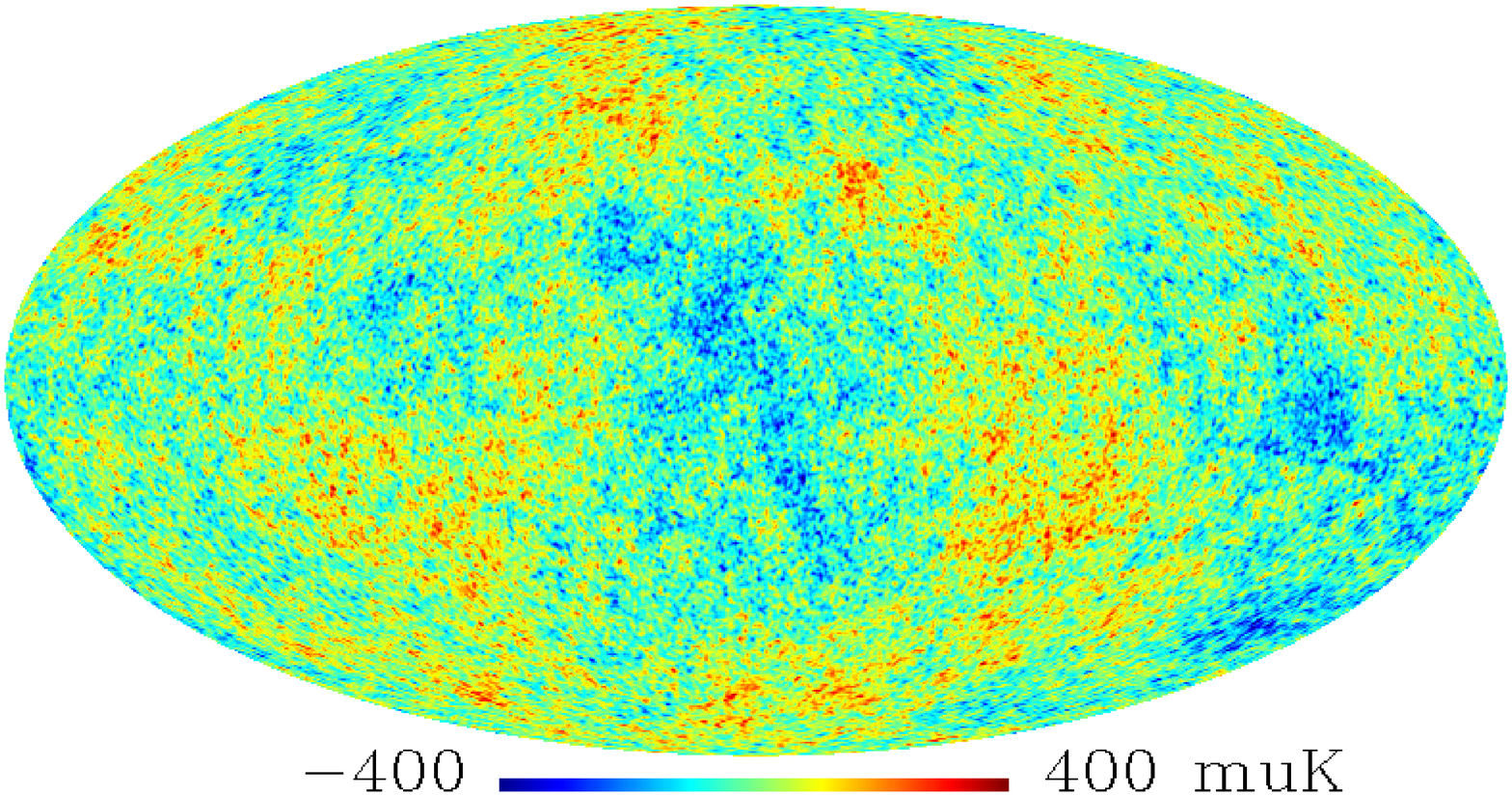}
\includegraphics[width=0.4\textwidth]{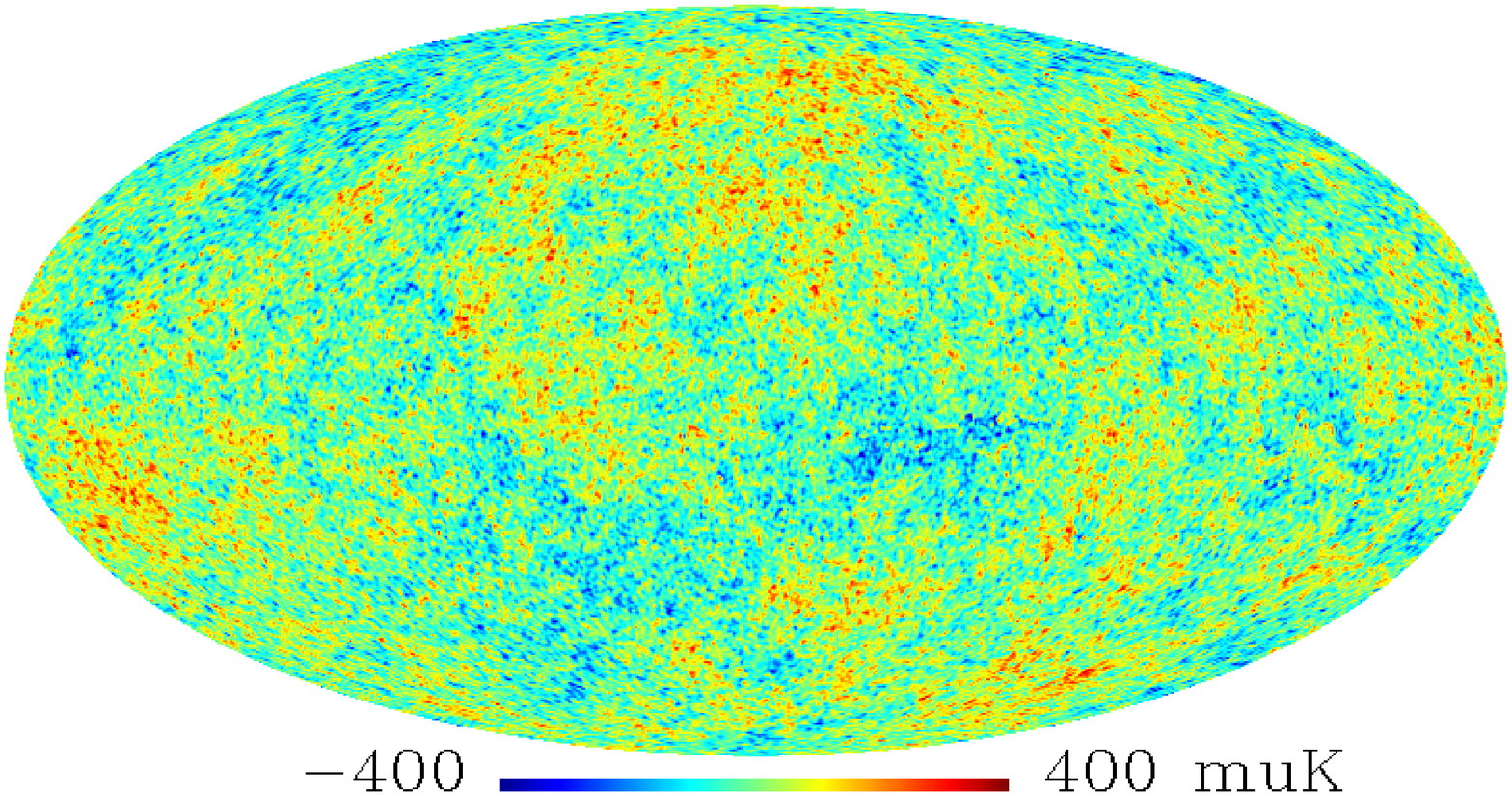}
\includegraphics[width=0.4\textwidth]{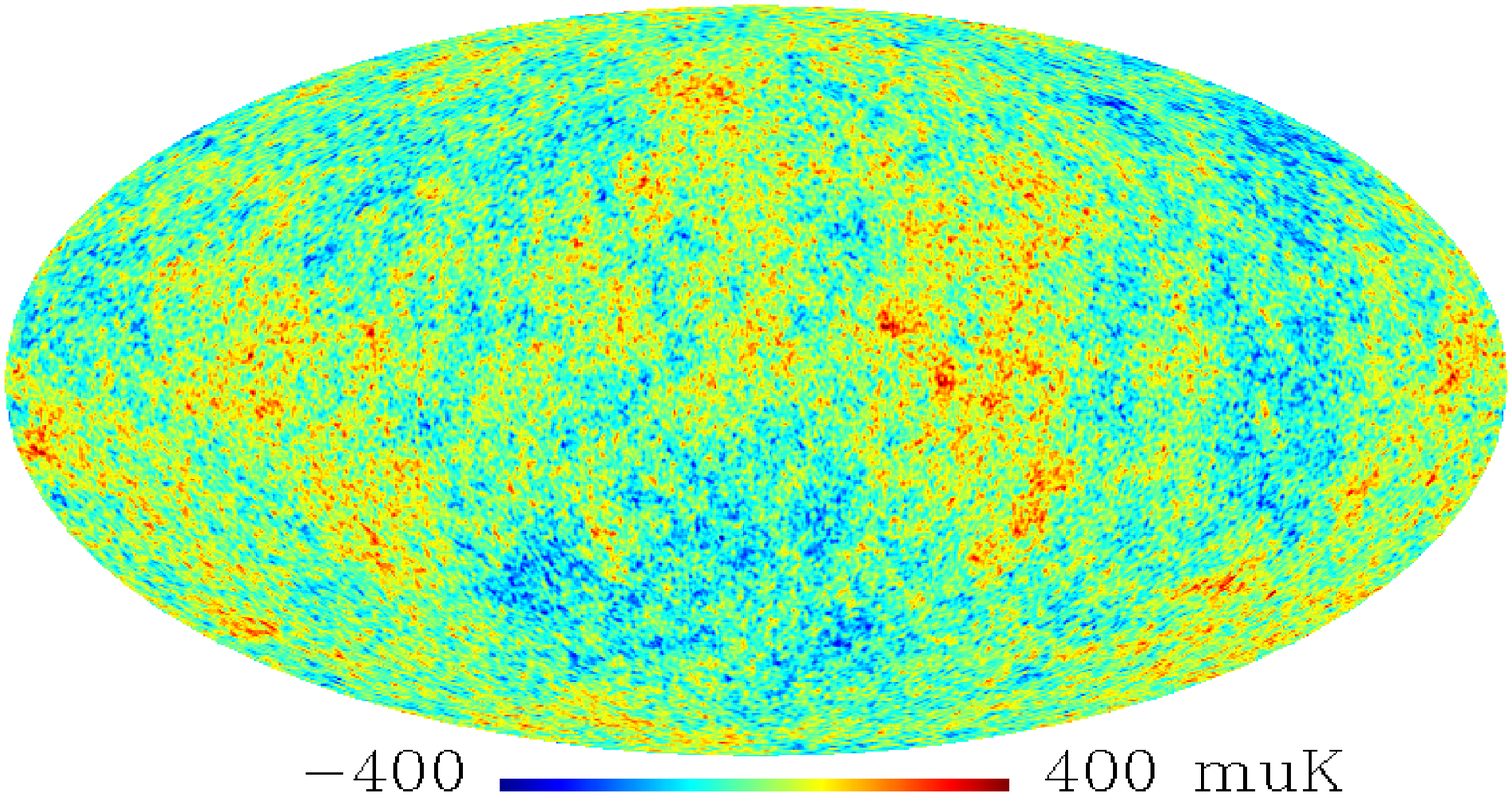}
\includegraphics[width=0.4\textwidth]{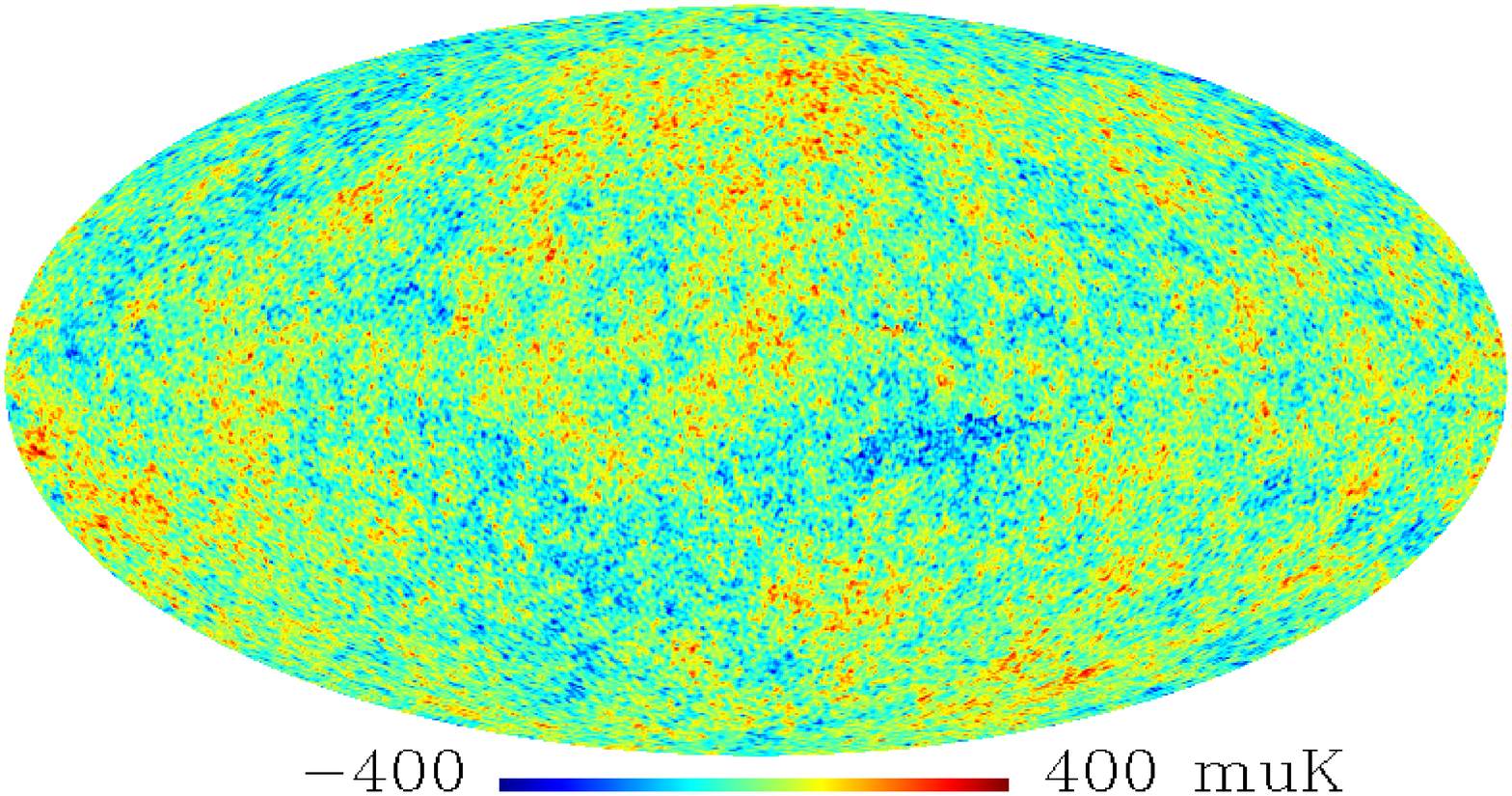}
\caption{the simulated input map with the KQ85 mask applied (top), unconstrained realization and 
constrained realization (the second and the third), another set of an unconstrained realization and a constrained realization (the fourth and bottom): the constrained realizations
are obtained by the Eq. \ref{cgh} with the preceding unconstrained realization respectively.}
\label{recon}
\end{figure}
\begin{figure}[!h]
\centering
\includegraphics[width=0.42\textwidth]{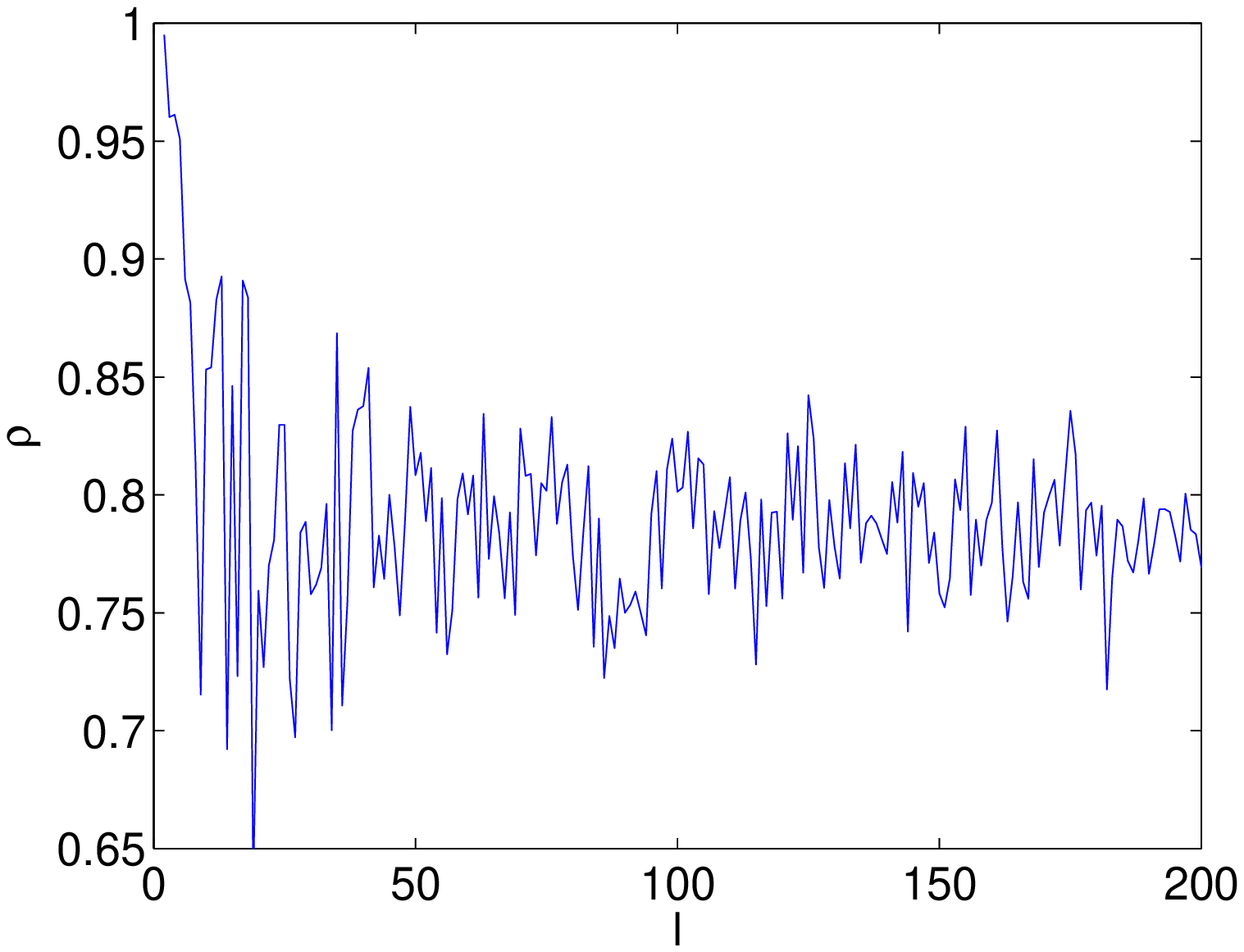}
\includegraphics[width=0.45\textwidth]{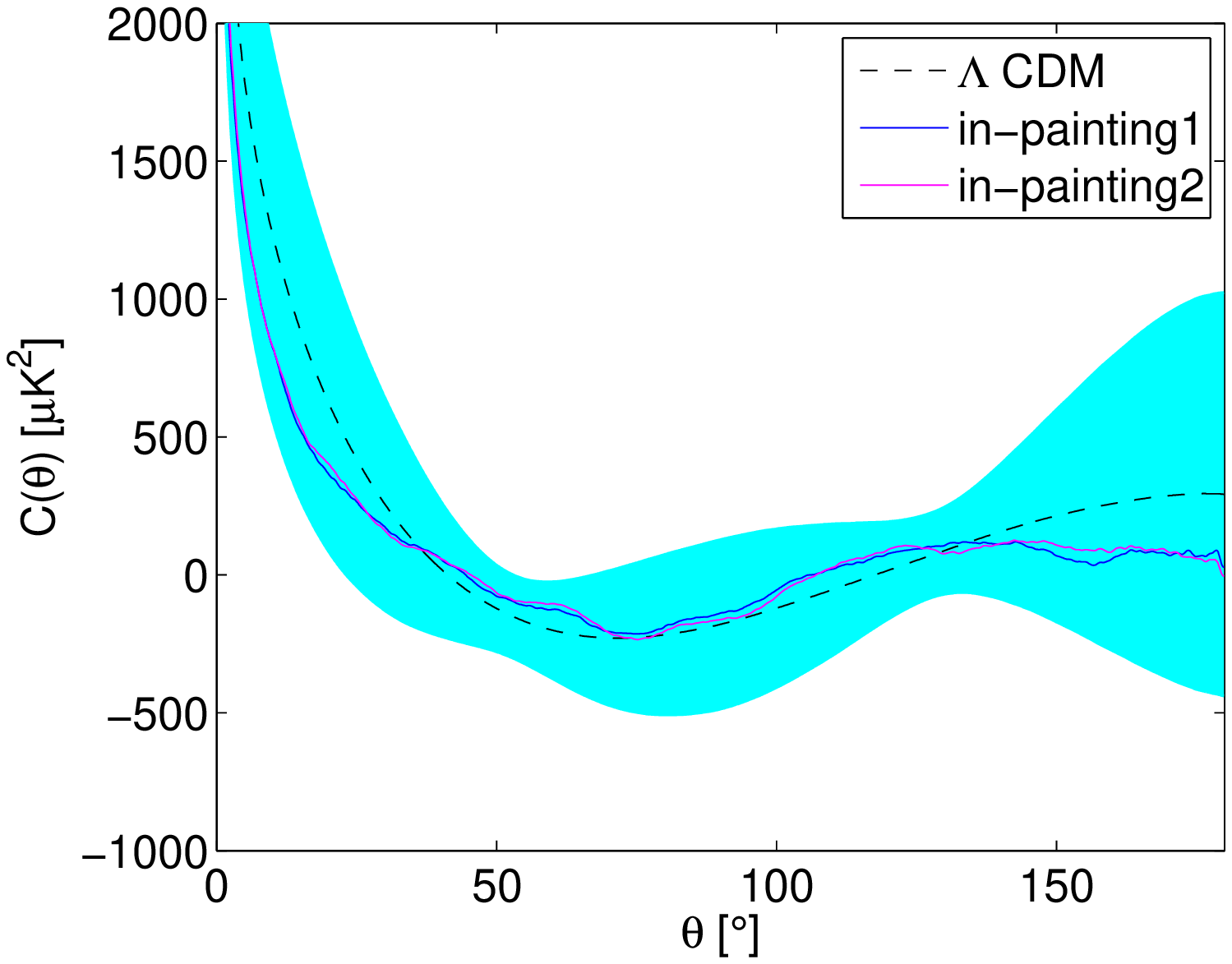}
\caption{Cross-correlation between in-painted maps for each multipole (top), angular autocorrelation of in-painted maps (bottom)}
\label{Cor}
\end{figure}

In order to test our method, we generated simulated CMB data, where we assumed the WMAP concordance $\Lambda$CDM model and the WMAP beam smoothing at V band \citep{WMAP7:Cosmology,WMAP7:basic_result}. We set the simulation to contain the multipoles up to 700, and produced it with the HEALPix pixellization Nside=512 \citep{HEALPix:Primer,HEALPix:framework}. 
We masked the simulated data by the WMAP KQ85, which admits pixel data of sky fractions 78\%.
At the top of Fig. \ref{recon}, we show the simulated map with the foreground mask applied.
We applied the procedure described in the previous section to the masked simulated data, and generated a whole-sky map.

In Fig. \ref{recon}, we show two constrained realizations obtained by Eq. \ref{cgh}, where we used different unconstrained realizations for the `mc' terms in Eq. \ref{cgh}. We like to stress that both of the results are equally likely, given the constraints. 
In Fig. \ref{Cor}, we show the correlation between two in-painted maps, which is computed for each multipole as follows: 
\begin{eqnarray}
\rho=\frac{\sum_m \mathrm{Re}[a_{1,lm}\,(a_{2,lm})^*]}{\sqrt{\sum_{m'} \left|a_{1,lm'}\right|^2\,\sum_{m''} \left|a_{2,lm''}\right|^2}},
\end{eqnarray}
where `1' and `2' in the subscripts denote two results respectively.
As shown in Fig. \ref{Cor}, we may see there is strong convergence at low multipoles, which also stays strong at higher multipoles. If we used a less conservative foreground mask, which, for instance, constitutes of large holes around Galactic plane like a Swiss cheese, the level of convergence will be even stronger. Nonetheless, given the random Gaussian nature of CMB anisotropy, it is not possible to reconstruct the exact realization, which happens to be our Universe. Therefore, our intention is filling-in the missing information in a way compliant with the expected statistical properties. 
However, we do not presumably expect the statistical properties of in-painted maps to be very close to the theoretical prediction, due to the statistical fluctuation associated with cosmic variance.

In order to see whether our in-painted maps, indeed, are consistent with the expected statistical properties, we estimated angular correlations, which are plotted in Fig. \ref{Cor}. 
In the same plot, we show the angular correlation of the WMAP concordance model \citep{WMAP7:Cosmology}, where the dotted line and shaded region denote the theoretical prediction and 2$\sigma$ ranges, as determined by Monte-Carlo simulations.
As shown in Fig. \ref{Cor}, we find the angular correlation of our in-painted maps are well inside the shaded region. We also find that the angular correlation of in-painted maps are similar to each other, which agrees with the strong convergence previously shown.

\section{Application to the WMAP data}
\label{wmap}

In order to reduce foregrounds, the WMAP team subtracted diffuse foregrounds by template fitting, and produced `foreground-reduced maps', which are available at Q, V and W band respectively \citep{WMAP7:fg}. Besides the foreground-reduced maps, there is the Internal Linear Combination (ILC) map, which is usually used without foreground masking. 
In spite of contamination from bright point sources and Galactic foregrounds, difficulty of investigating lowest multipoles on a masked sky data made the whole-sky ILC widely used for the investigation of CMB data anomalies \citep{Tegmark:Alignment,Multipole_Vector1,Multipole_Vector2,Multipole_Vector3,Axis_Evil,Axis_Evil2,Axis_Evil3,Universe_odd,Phase_correlation,Chiang_NG,odd,odd_origin,lowl_anomalies,lowl_bias,odd_phase,CMB_octupole,Parity_Dipole}.
However, foreground-reduced maps with foreground mask is more reliable and contains less foreground contamination than the ILC map. Therefore, it is worth to investigate the anomaly at lowest multipoles, using the foreground-reduced maps.

Using our method, we in-painted the masked foreground-reduced maps.
The in-painting process is the same with the one described in the previous sections, except that $C_l$ in Eq. \ref{FFCl} should be replaced by $C_l+N_l$ with $N_l$ being the power spectrum of instrument noise. 
We simulated instrument noise by the WMAP noise model $\sigma_0/\sqrt{N_{\mathrm{obs}}}$ \citep{WMAP7:basic_result}, and estimated the noise power spectrum $N_l$ from 1000 simulated noise maps, where $N_{\mathrm{obs}}$ is the number of observations for a pixel and $\sigma_0$ is 2197, 3137 and 6549 [$\mu$K] for Q, V and W band respectively.
For the foreground mask, we used the WMAP team's KQ85 mask.

\begin{figure}
\centering
\includegraphics[width=0.4\textwidth]{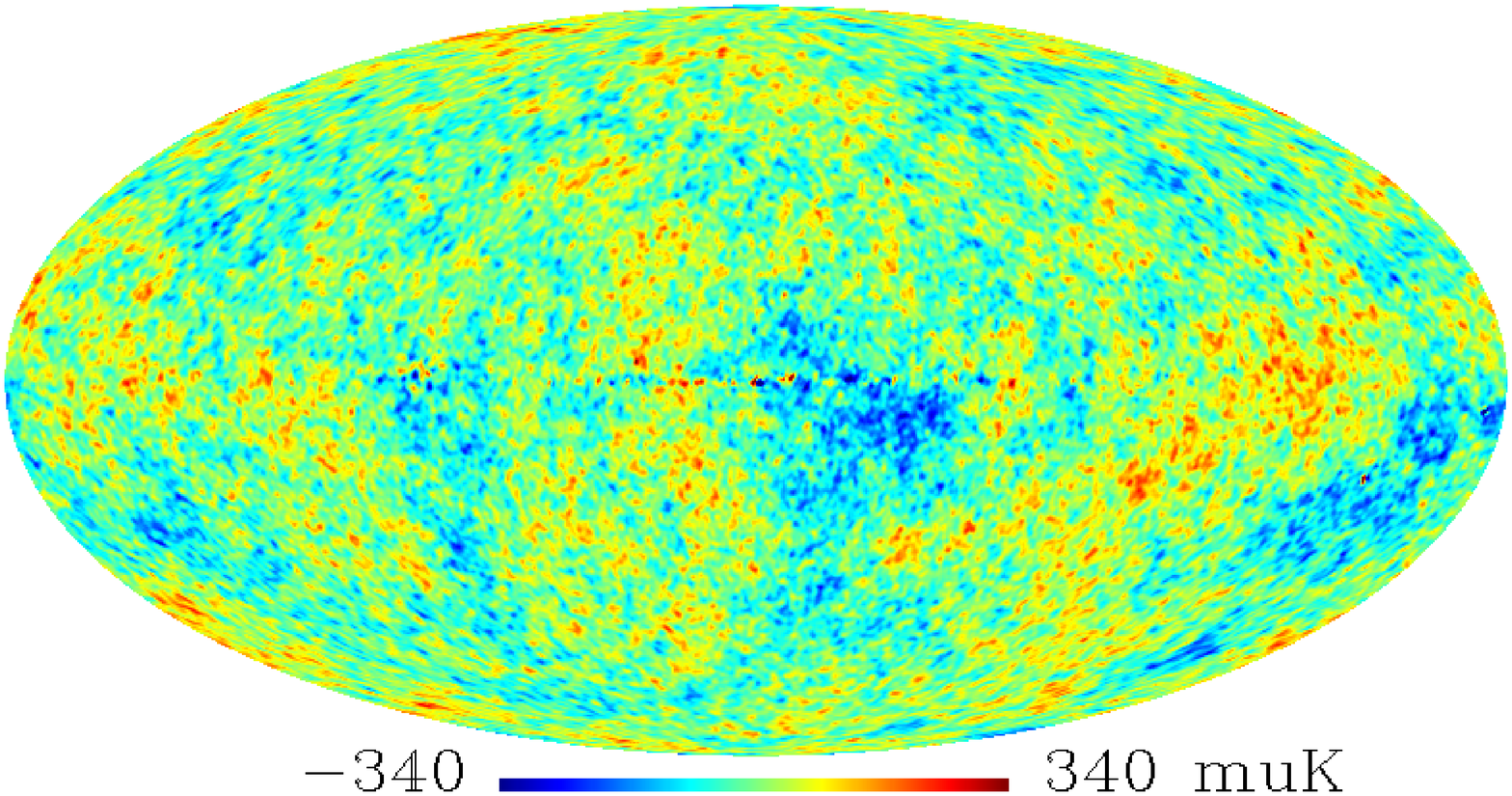}
\includegraphics[width=0.4\textwidth]{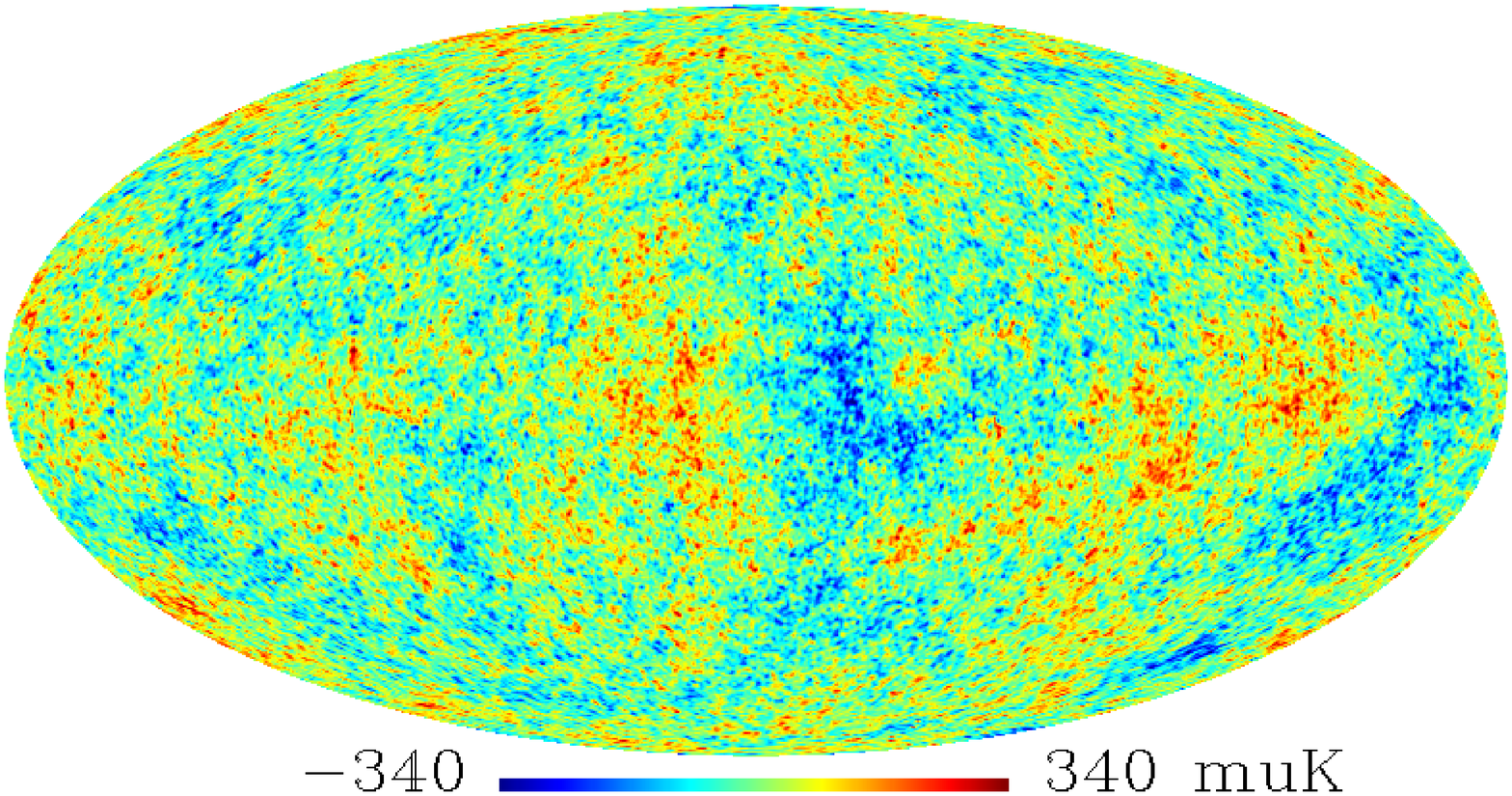}
\includegraphics[width=0.4\textwidth]{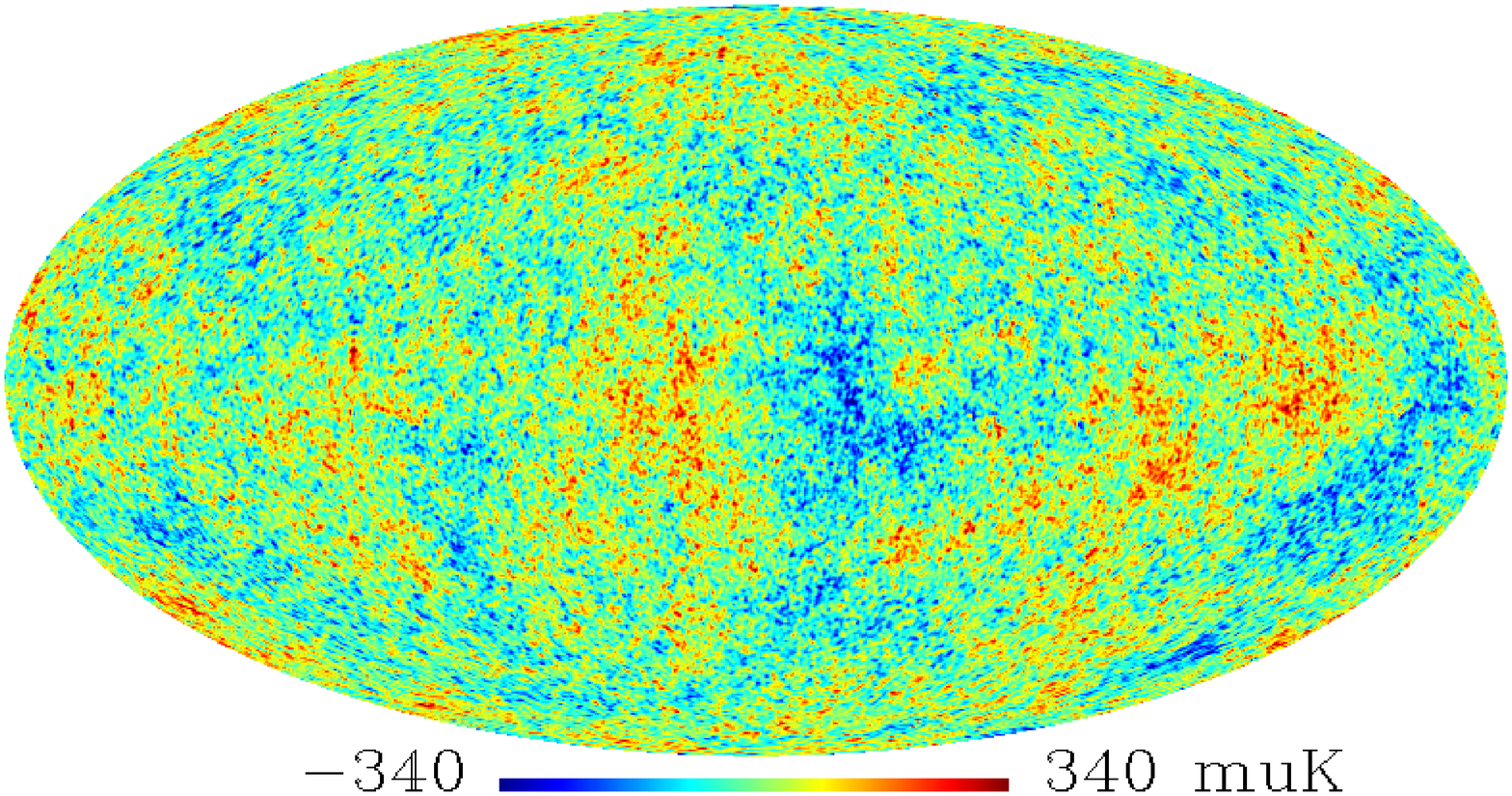}
\includegraphics[width=0.4\textwidth]{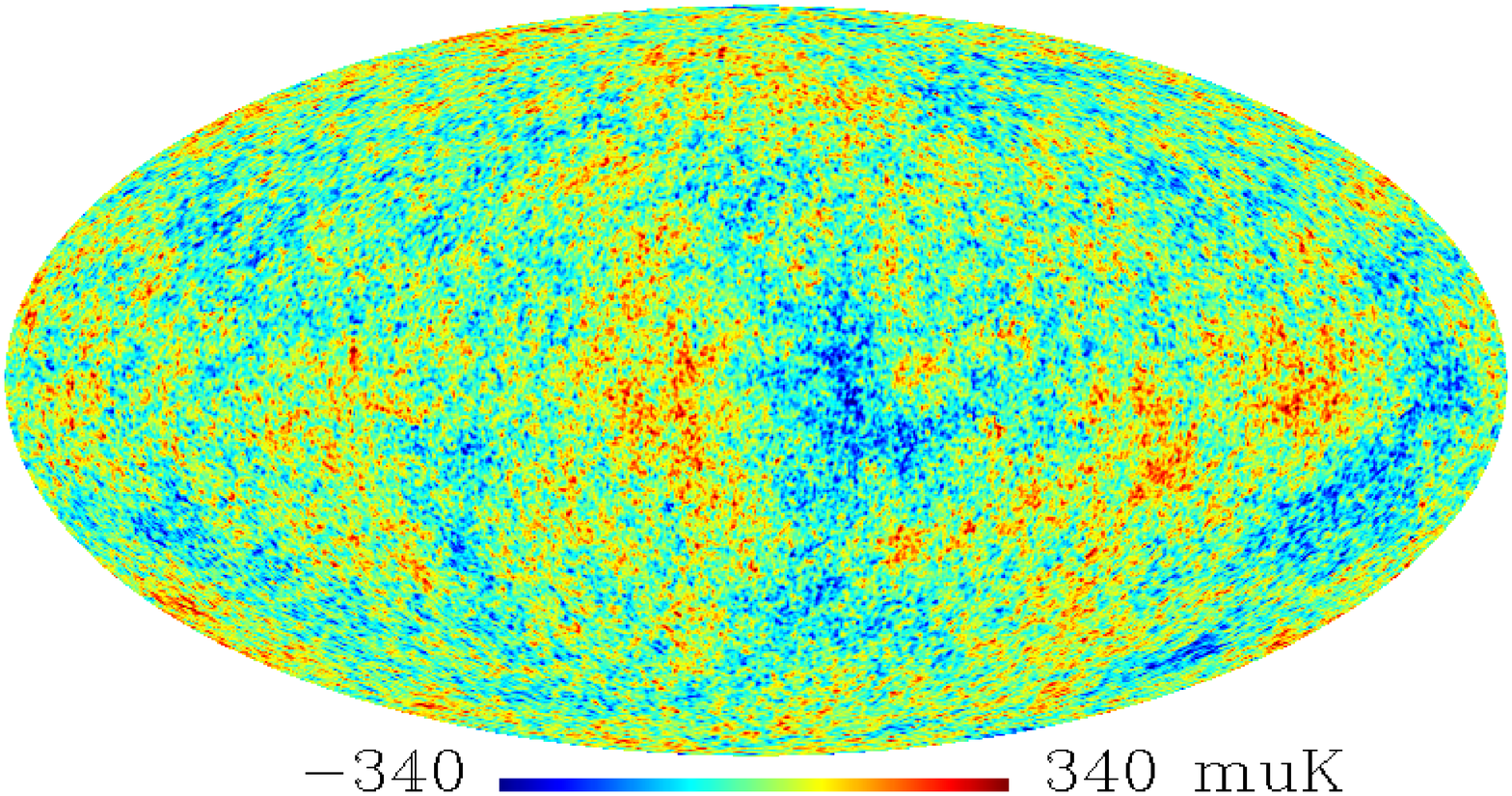}
\caption{the ILC map (top), in-painted maps of Q, V, W band (from the second to the last)}
\label{WMAP}
\end{figure}
\begin{figure}
\centering
\includegraphics[width=0.23\textwidth]{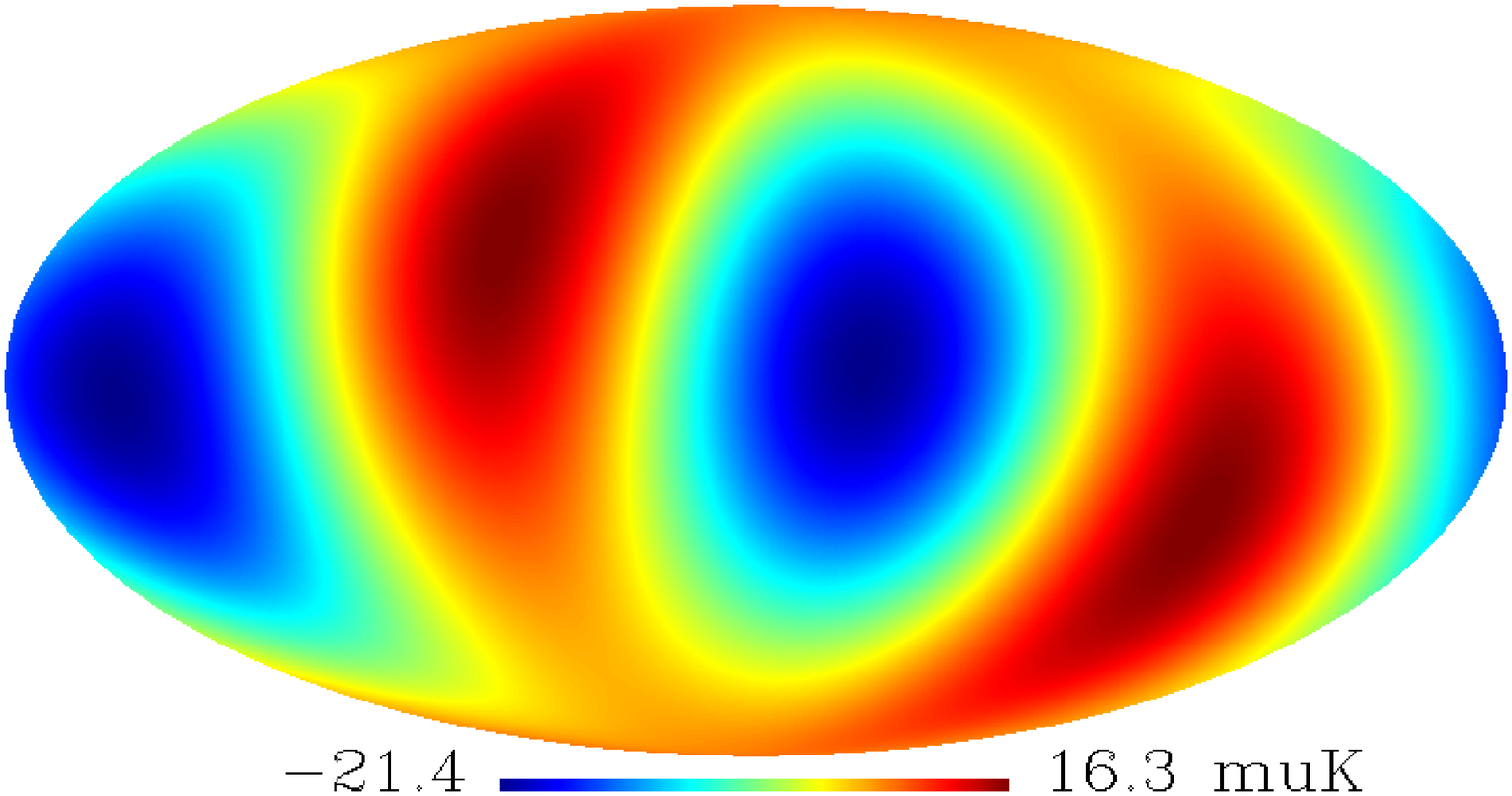}
\includegraphics[width=0.23\textwidth]{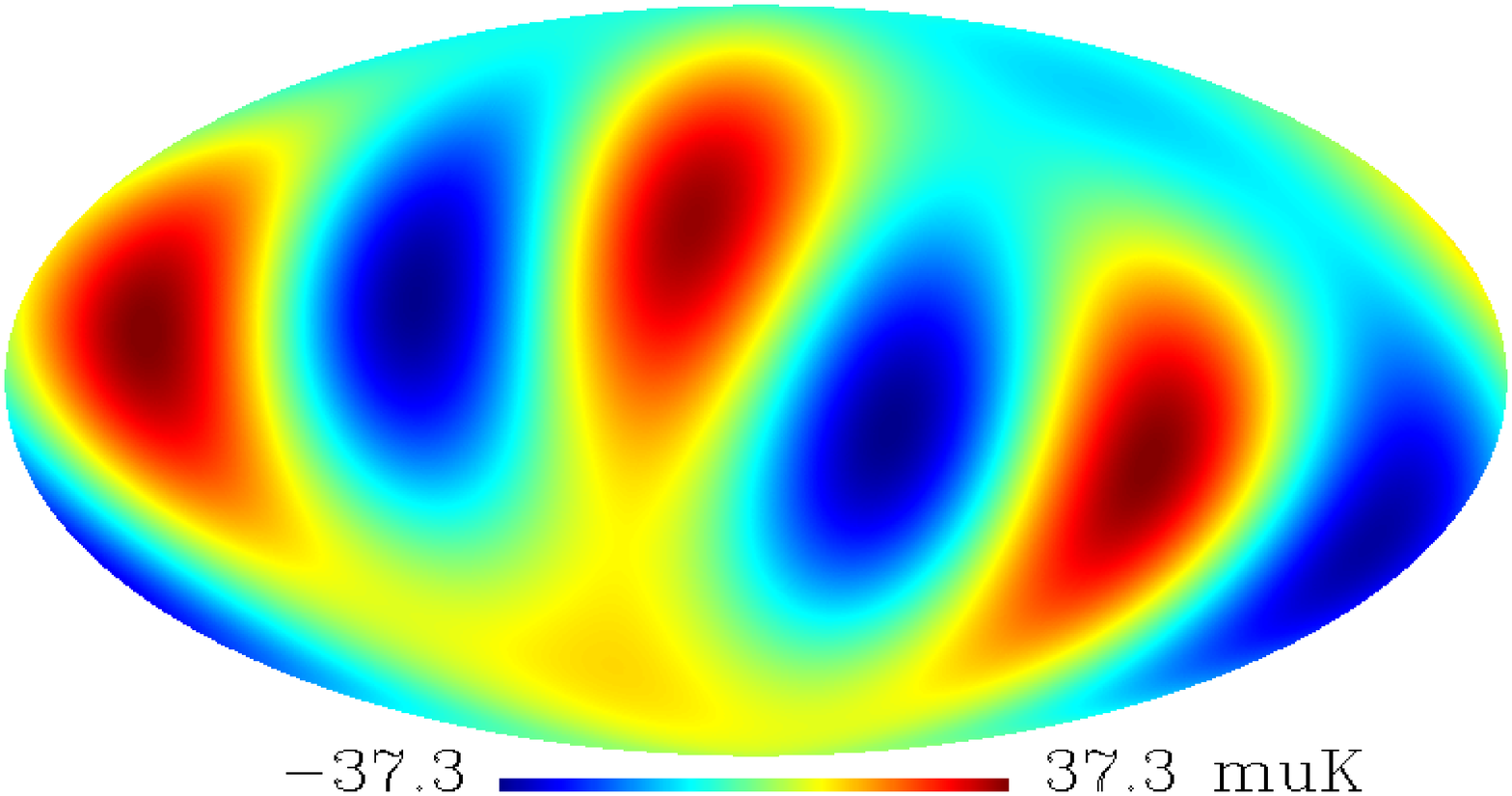}
\includegraphics[width=0.23\textwidth]{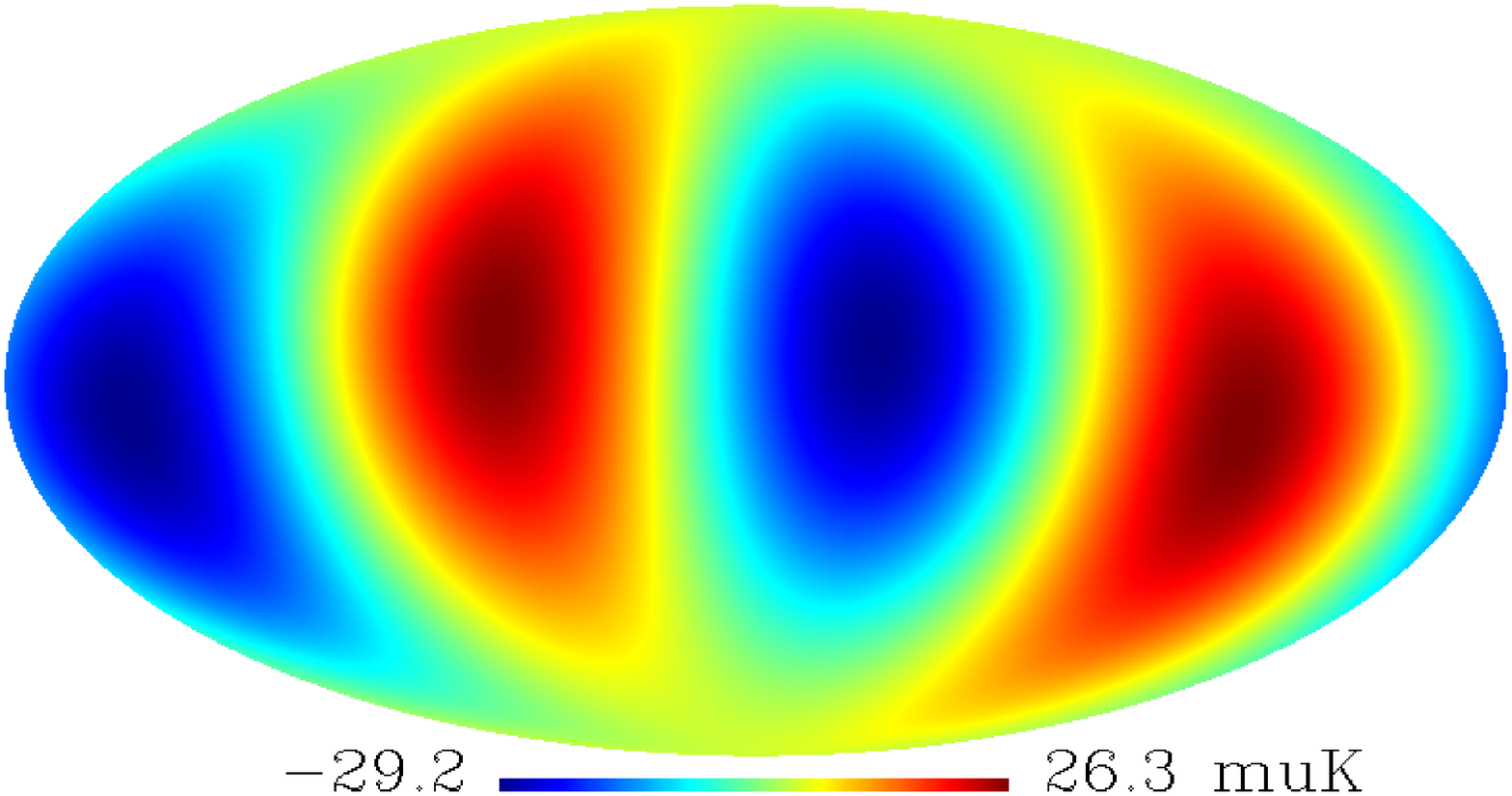}
\includegraphics[width=0.23\textwidth]{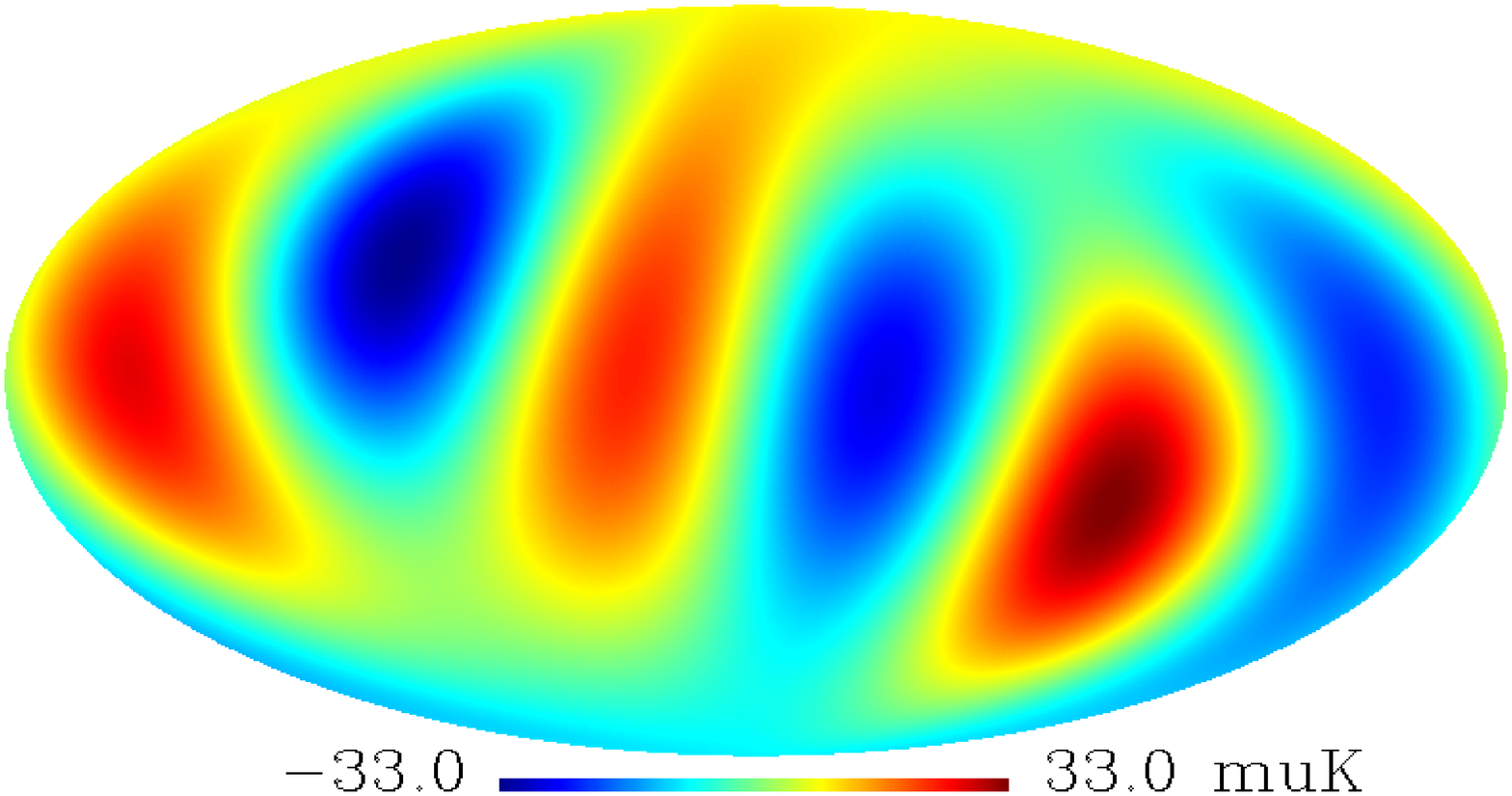}
\includegraphics[width=0.23\textwidth]{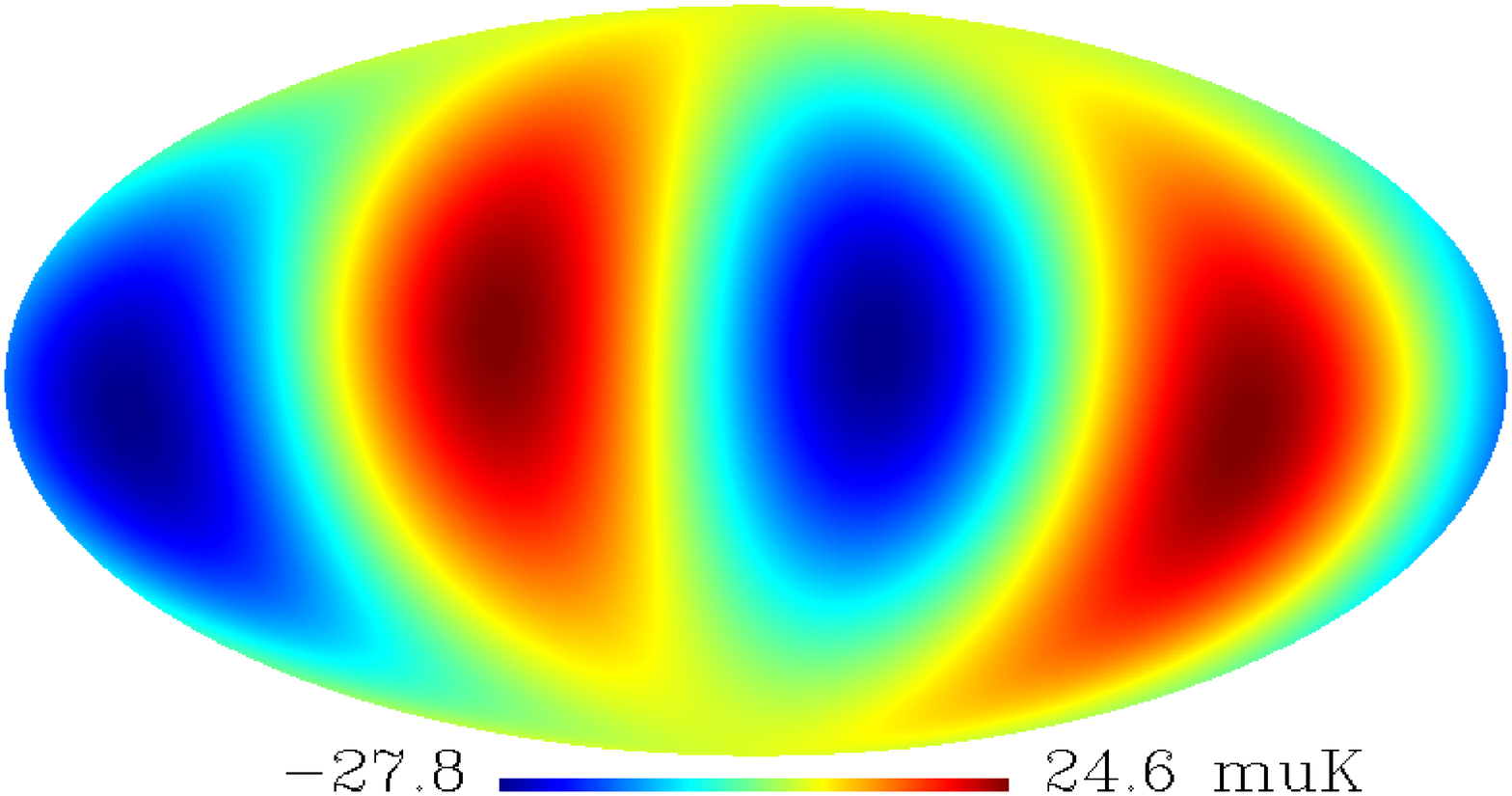}
\includegraphics[width=0.23\textwidth]{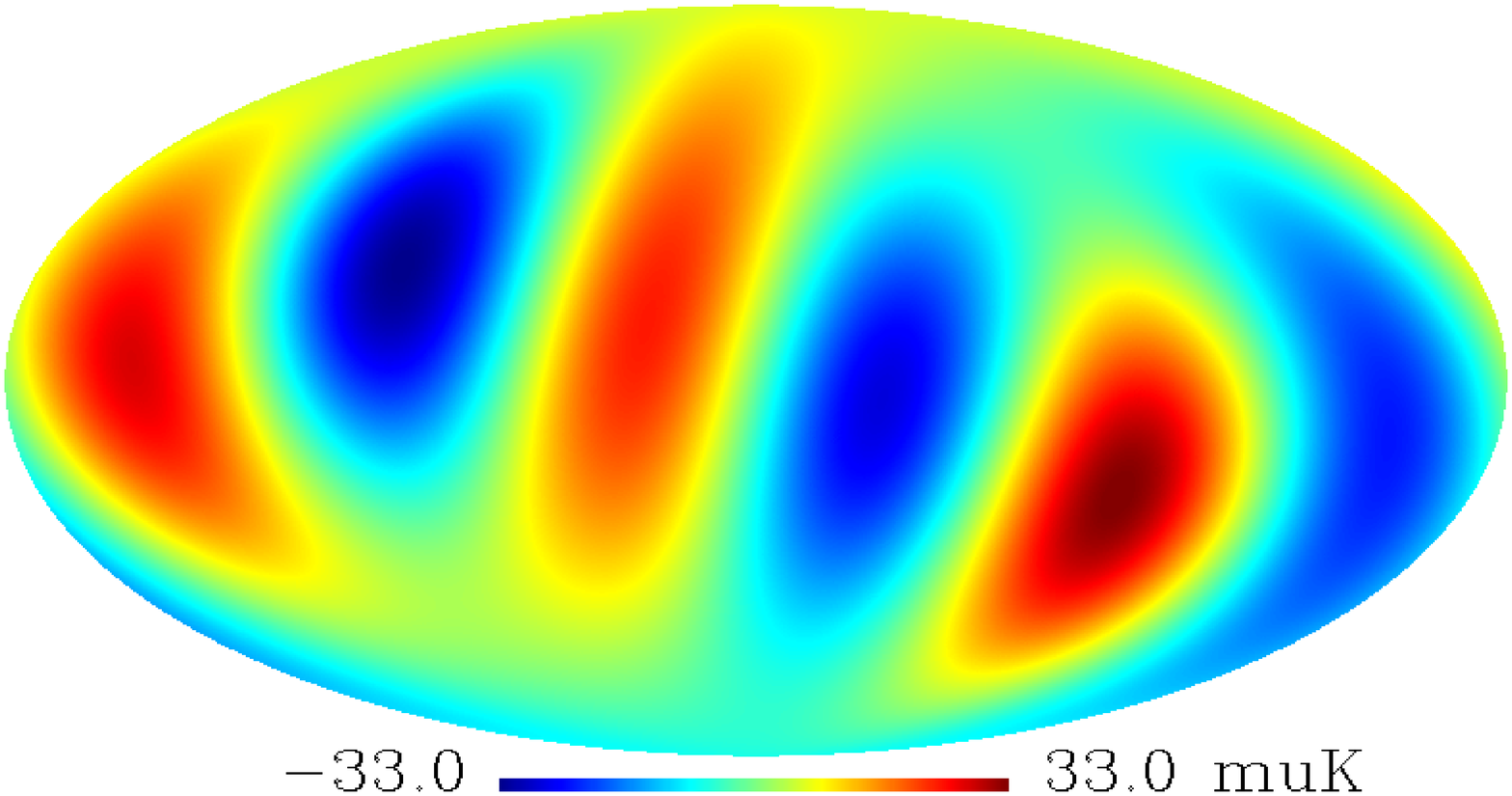}
\includegraphics[width=0.23\textwidth]{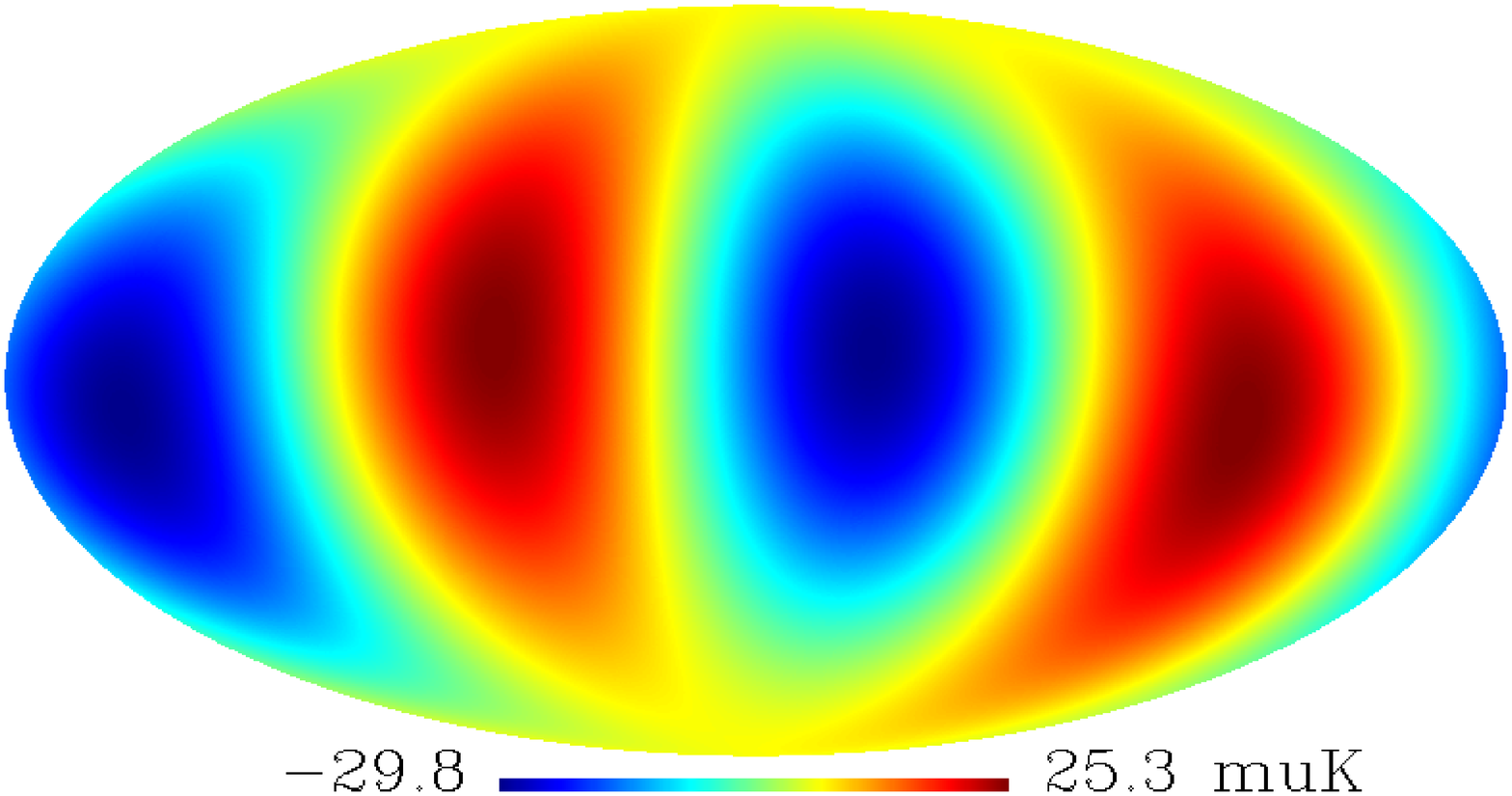}
\includegraphics[width=0.23\textwidth]{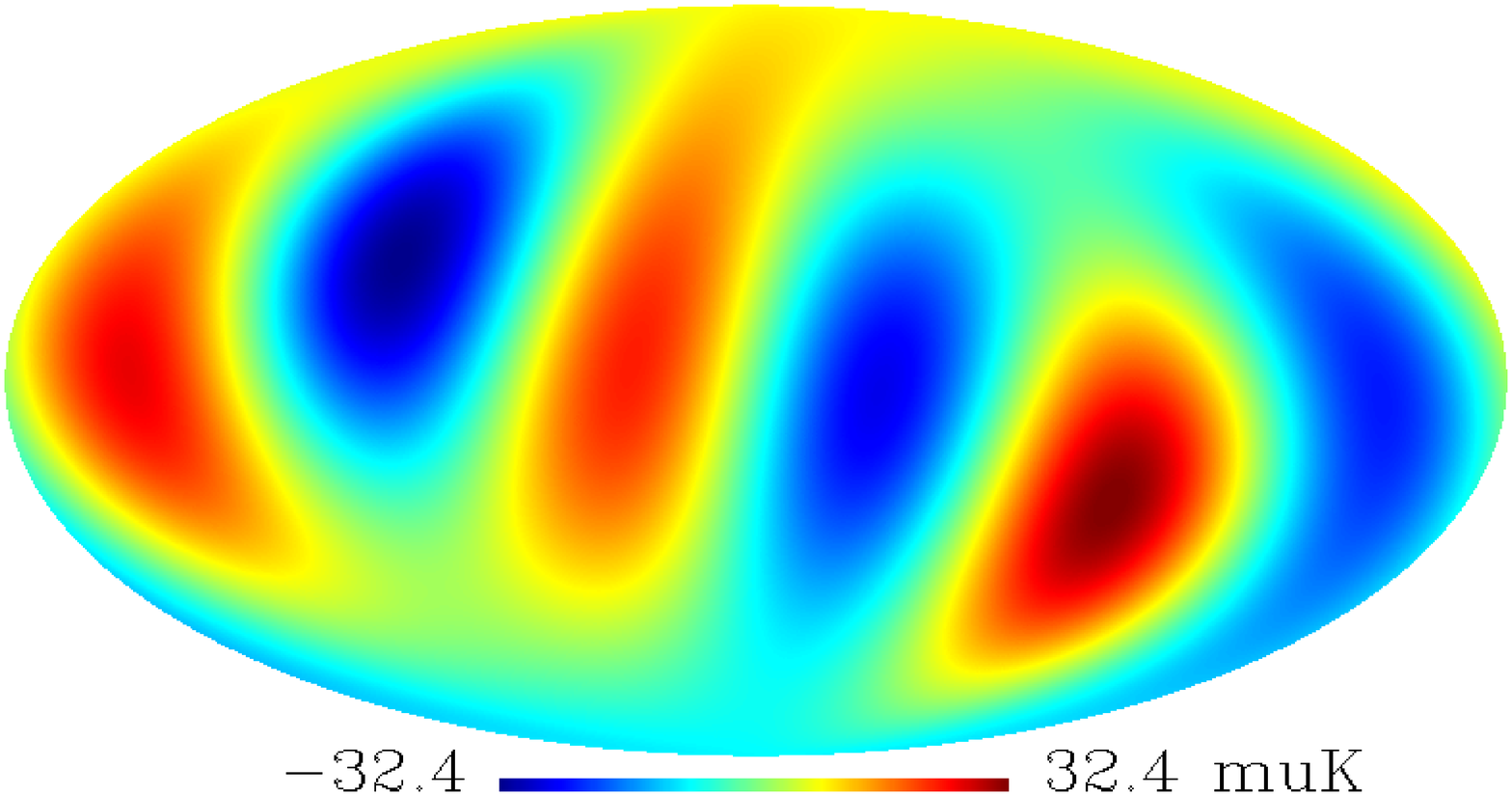}
\caption{the quadrupole components (left) and octupole (right) of the ILC map (top), in-painted maps of Q, V, W band (from the second to the last)}
\label{l23}
\end{figure}
In Fig. \ref{WMAP}, we show the whole-sky ILC map and our in-painted maps. Note that in-painted maps have the beam smoothing of the original maps, which are distinct at each band. In particular, the beam smoothing of the ILC map, which corresponds to FWHM=$1^\circ$, differs most significantly from the others.
From the in-painted maps, we estimated the quadrupole and octupole components, which are shown in Fig. \ref{l23}.
For comparison, we show the quadrupole and octupole components of the WMAP ILC map.
It is interesting to notice that the amplitudes of quadrupole and the octupole anisotropy are greater and smaller than the ILC map respectively, which will alleviate the anomaly of the low quadrupole power and the parity asymmetry \citep{Tegmark:Alignment,odd,odd_origin,odd_bolpol}. Further investigation on this issue is deferred to separate publications. 

Using the quadrupole and octupole anisotropy of in-painted maps, we investigated the anomalous alignment between the quadrupole and octupole, which are found in the multipole vector analysis of the ILC map \citep{Multipole_Vector1,Multipole_Vector2,Multipole_Vector3}.
In the original study by  \citep{Multipole_Vector1,Multipole_Vector2,Multipole_Vector3}, three dot products $D_1$, $D_2$, and $D_3$ were estimated, where the most anomalous alignment is associated with $D_1$.  
\begin{figure}
\centering
\includegraphics[width=0.45\textwidth]{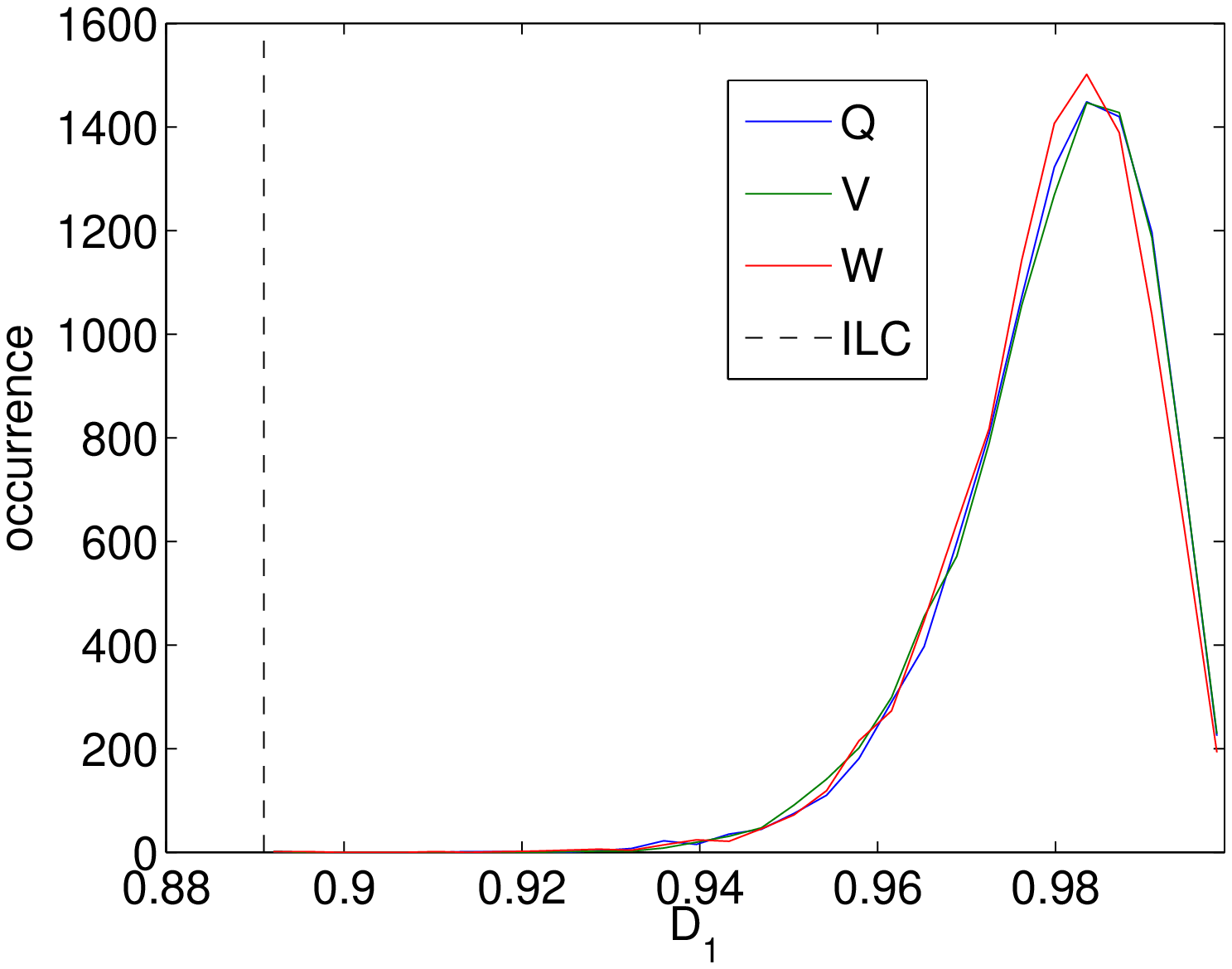}
\includegraphics[width=0.45\textwidth]{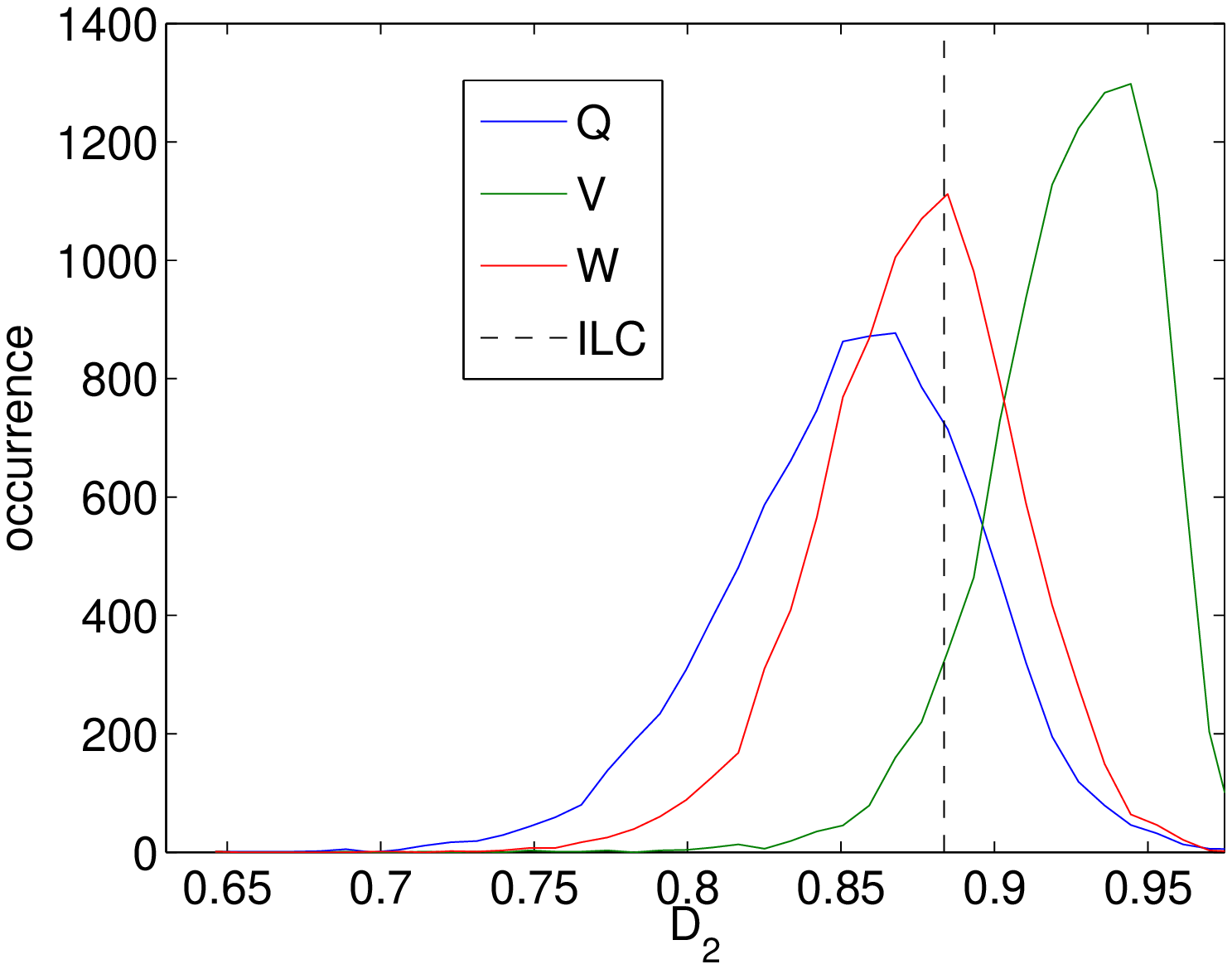}
\includegraphics[width=0.45\textwidth]{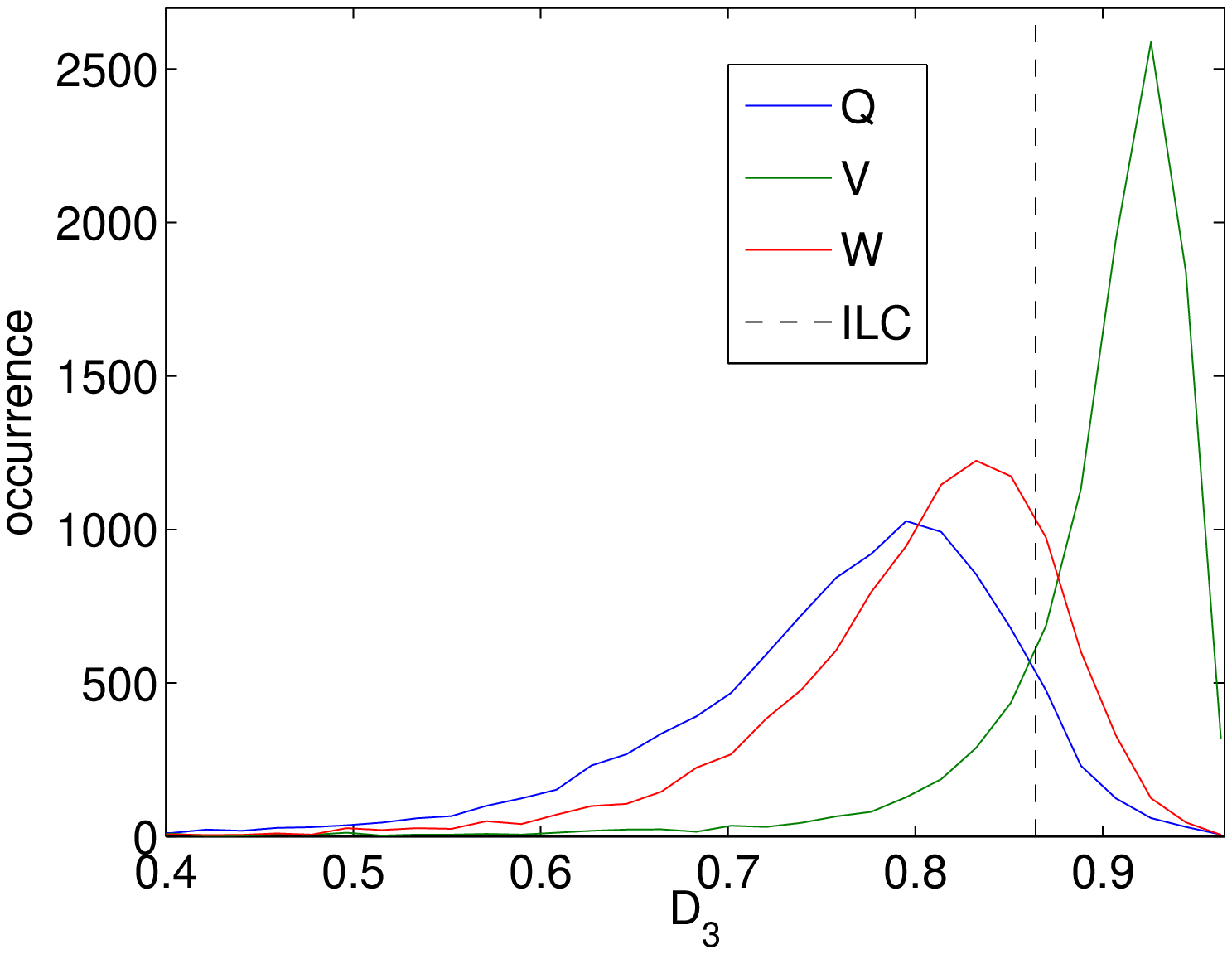}
\caption{the distribution of the multipole vector alignment between the quadrupole and octupole components of in-painted maps: the alignment of the whole-sky ILC is marked as dashed lines for comparison.}
\label{alignment}
\end{figure}
Since our in-painting method is statistical, we generated 10,000 in-painted maps for Q, V and W bands respectively.
From each in-painted map, we computed the alignment between multipole vectors, which are quantified by three dot products $D_1$, $D_2$, and $D_3$, where a higher value of a dot product correspond to higher alignment \citep{Multipole_Vector1,Multipole_Vector2,Multipole_Vector3}. 
In Figure 5, we show the distribution of the dot product values. For $D_1$, we find all in-painted maps, except for one in-painted map of W band, have even higher alignment than the ILC map. 
It is also interesting to note that $D_2$ and $D_3$ values of the V band map are much higher than the other bands, even though the V band map is expected to contain least foreground contamination.
Given our result, we find it difficult to attribute the anomalous alignment to the residual foregrounds.
Previously, \cite{PE_low,PE_mvec} investigated anisotropy at lowest multipoles, by applying the power equalization filter to the cut-sky foreground-reduced maps.
Our result is consistent with their finding that the alignment anomaly is robust with respect to the frequency and sky cut \citep{PE_low,PE_mvec}.

\section{Discussion}
\label{discussion}
Foreground masking leads to large cuts more or less parallel to the Galactic equator and numerous holes in the CMB map.
Therefore, there have been a lot of effort on in-painting of CMB sky map.
Though there have been well-established methods on constrained Monte-Carlo simulation for Gaussian fields, the prohibitive computational cost makes it unfeasible for the WMAP or Planck data. In this work, we implemented in-painting in harmonic space, which the computational load may be greatly reduced with good approximation. 
We applied our method to simulated data and the WMAP data. 
It should be kept in mind that the method and the result presented in this work are valid only to the extent our early Universe is Gaussian and statistically isotropic.
In the result with the simulated data, we found the the angular correlation of the in-paint maps are in good agreement with the assumed model.
Using the in-painted maps of WMAP data, we investigated the anomalous alignment between the quadrupole and octupole components, which are originally found in the multipole vector analysis of the WMAP whole-sky ILC map. From the distribution of $D_1$ values, we find the alignment in the foreground-reduced map is even higher than that of the ILC map.
It is interesting to notice that V band maps show rather higher alignment than other bands, despite the expectation of the V band map being cleanest.
Therefore, we find it hard to attribute the alignment to residual foregrounds.
The alignment anomaly, including other anomalies, deserve more rigorous investigations combined with our in-painting method.
However, in this letter, we contend ourselves with demonstrating our method, and defer more rigorous investigation to future publications.
For the Planck data analysis and future missions, we believe our method will be of great use.

\section{Acknowledgments}
We are grateful to an anonymous referee for thorough reading and comments, which greatly helped us to improve the clarity of this work.
We acknowledge the use of the Legacy Archive for Microwave Background Data Analysis (LAMBDA).
Our data analysis made the use of HEALPix \citep{HEALPix:Primer,HEALPix:framework} and SpICE \citep{spice2,spice1}.  
This work is supported in part by Danmarks Grundforskningsfond, which allowed the establishment of the Danish Discovery Center.

\bibliographystyle{plainnat}
\bibliography{/home/tac/jkim/Documents/bibliography}
\end{document}